# Observation of stable HO$_4^+$ and DO$_4^+$ ions from ion-molecule reactions in helium nanodroplets

Michael Renzler,[1] Stefan Ralser,[1] Lorenz Kranabetter,[1] Erik Barwa,[1] Paul Scheier[1,*] and Andrew M. Ellis[2,*]

[1] Institut für Ionenphysik und Angewandte Physik, Universität Innsbruck, Technikerstr. 25, A-6020 Innsbruck, Austria

[2] Department of Chemistry, University of Leicester, University Road, Leicester, LE1 7RH, UK

Email: Paul.Scheier@uibk.ac.at; andrew.ellis@le.ac.uk

___________________________________________________________

**Abstract**

Ion-molecule reactions between clusters of $H_2/D_2$ and $O_2$ in liquid helium nanodroplets were initiated by electron-induced ionization (at 70 eV). Reaction products were detected by mass spectrometry and can be explained by a primary reaction channel involving proton transfer from $H_3^+$ or $H_3^+(H_2)_n$ clusters and their deuterated equivalents. Very little $HO_2^+$ is seen from the reaction of $H_3^+$ with $O_2$, which is attributed to an efficient secondary reaction between $HO_2^+$ and $H_2$. On the other hand $HO_4^+$ is the most abundant product from the reaction of $H_3^+$ with oxygen dimer, $(O_2)_2$. The experimental data suggest that $HO_4^+$ is a particularly stable ion and this is consistent with recent theoretical studies of this ion

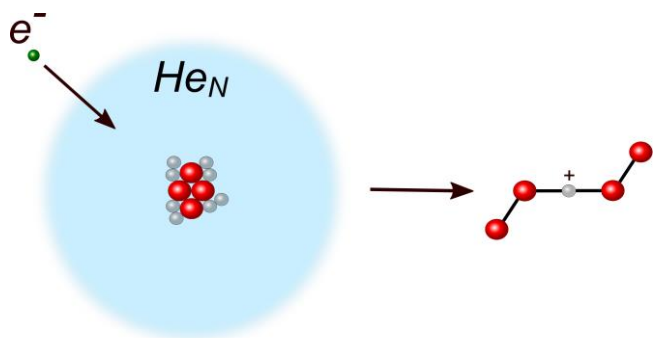

*Novelty of work*: Experimental observation of the enhanced stability of the $HO_4^+$ ion has been made in a study using helium nanodroplets.

The lack of a permanent electric dipole moment makes the detection of $O_2$ in astrophysical settings a significant challenge. It is only very recently that direct observations of $O_2$ have been made in the interstellar medium (ISM) through the detection of weak rotational lines driven by magnetic dipole transitions in the millimeter region of the spectrum.[1-5] Nevertheless, there are significant differences between predicted and observed abundances of $O_2$ in the ISM[6] and therefore an alternative means of quantifying $O_2$ would be valuable.

Many years ago it was suggested that detection of protonated molecules might provide an indirect method for quantifying homonuclear diatomics such as $N_2$, $O_2$ and $C_2$.[7] When protonated these molecules are expected to possess a substantial electric dipole moment and should therefore be easy to detect by rotational or vibrational spectroscopy if they are reasonably abundant. The most likely source of protons in the ISM is $H_3^+$ and laboratory studies have shown that this ion protonates $N_2$ in a fast exothermic reaction.[8] The resulting $N_2H^+$ ion is a tracer molecule for $N_2$ and has indeed been used to determine the abundance of $N_2$ in dense interstellar clouds.[7] The possibility therefore exists for using $O_2H^+$ in a similar manner. Unfortunately, $O_2$ possesses a lower proton affinity than $N_2$ and a detailed analysis has shown that the reaction

$$H_3^+ + O_2 \rightarrow HO_2^+ + H_2 \qquad (1)$$

is slightly endothermic.[9] Although endothermic by only $50 \pm 9$ cm$^{-1}$ at 0 K, this is enough to provide a strong obstacle to this reaction at the low temperatures found in many astronomical environments and means that $HO_2^+$ is of little or no value as a tracer for oxygen in the ISM.

Very recently, it has been suggested that $HO_4^+$ might be used as an alternative to $HO_2^+$ as a tracer molecule for $O_2$.[10,11] The basic assumption here is of reaction of the $(O_2)_2$ dimer, instead of $O_2$, with $H_3^+$, although as discussed later it is questionable whether the dimer is significant in the ISM. The proton affinity of the oxygen dimer, $(O_2)_2$, has not been measured but one would expect a higher value than for monomeric $O_2$ because the proton can be shared between two molecules. This presumption is confirmed by *ab initio* calculations, which predict a proton affinity for the dimer which is 0.84 eV higher than that of $O_2$.[11] This will now make the $O_4$ equivalent of reaction (1) substantially exothermic and, given that exothermic ion-molecule reactions usually have no activation energy, proton transfer from $H_3^+$ is likely to approach the diffusion limited rate. Moreover, $HO_4^+$ is an ion which has received little prior study by theory or experiment. The first and only previous experimental observation of $H(O_2)_n^+$ ions was derived from mass spectrometric work in the gas phase using a high pressure ion source.[12] On the basis of observed abundances of ions as a function of temperature an $O_2$-$O_2H^+$ binding energy of 86.1 kJ mol$^{-1}$ was deduced. This value is more than an order of magnitude greater than the binding energy of the neutral $(O_2)_2$ dimer[13] and shows that the proton induces quite strong binding between the two $O_2$ molecules. Two recent and related *ab initio* studies have suggested that $HO_4^+$ is a rather interesting molecule with a Zundel-like structure reminiscent of the protonated water dimer, $H_5O_2^+$.[10,11] According to these calculations the most stable structure is a trans isomer possessing $C_{2h}$ point group symmetry.

In this study we have explored the ion-molecule chemistry between hydrogen and oxygen in helium nanodroplets and we report specifically on the observation of ions of the type $H_mO_x^+$, where $x$ is even and includes $HO_4^+$. Helium droplets provide a very low temperature (0.37 K)[14] and gas-like environment in which to initiate ion-molecule reactions. We have performed experiments with both $H_2$ and $D_2$, where the latter makes it easier to rule out contributions from ions such as $H_2O^+$. Despite this potential complication for $H_2$ we nevertheless see similar results for $H_2$ and $D_2$. However, for the sake of simplicity we present data only from the $D_2$ experiments here. Oxygen and deuterium molecules were added sequentially to helium nanodroplets having a mean size in the region of $3 \times 10^5$ helium atoms. At the partial pressures employed we estimate an average pick-up of 13 $O_2$ and 14 $D_2$ molecules per droplet, although a broad distribution of mixed cluster sizes is expected on account of the stochastic nature of the pick-up process. The droplets were then subjected to bombardment by electrons at energies of 70 eV and any resulting ions were detected by a high resolution reflectron time-of-flight mass spectrometer. Full details of the apparatus can be found elsewhere.[15]

The ionization of pure hydrogen clusters in helium nanodroplets has been studied previously by Jaksch et al.[16] As well as seeing abundant $H_n^+$ clusters with odd $n$, clusters with even $n$ were also detected. The preferential formation of odd $n$ ions is a consequence of the facile reaction of $H_2^+$ with $H_2$ to give $H_3^+ + H$. The resulting $H_3^+$ can then combine with one or more $H_2$ molecules to give $H_n^+$ ions with odd $n$ and these are the dominant species observed. We take these ions as the starting point for the discussion here and consider what happens when oxygen is also added to the helium droplets.

To demonstrate the quality of the mass spectrometric data, Figure 1 shows part of the mass spectrum recorded for a $D_2/O_2$ mixture in helium nanodroplets. Figure 2 shows the yields of $D_mO_x^+$ ions as a function of $m$ for $x = 2, 4, 6$ and 12. In all four cases we see an odd-even intensity alternation in $m$, with the odd $m$ ions having a greater abundance than those with even $m$. This is consistent with the known findings for pure $H_2$ and $D_2$ in helium droplets and suggests that $D_3^+$ and its clusters are generated by the route indicated in the previous paragraph. Although molecular oxygen has a lower ionization energy than hydrogen, we expect the hydrogen to be ionized initially because it is added second to the helium droplets and therefore will be the first to come into contact with $He^+$ or $He^*$. Further evidence in favour of initial ionization of hydrogen comes from the known ionization behavior of molecular oxygen clusters, $(O_2)_n$. Electron ionization of these clusters preferentially produces ions with even $n$.[17] The mass spectrum in Figure 1 illustrates the predominance of even oxygen cluster ions in our observations but these even oxygen ions are expected to be unreactive with $H_2$ and $D_2$.[18] Consequently, our observations strongly suggest that the ion-molecule chemistry is initiated by reactions of cationic hydrogen and deuterium clusters.

We first consider detected ions containing $O_4$. Here the most abundant ion is $DO_4^+$. At a slightly lower abundance is $D_2O_4^+$, but thereafter the ion yields drop significantly and the plot relaxes into a simple odd-even oscillation pattern. The presence of significant excesses of only $DO_4^+$ and $D_2O_4^+$ ions allows us to rule out simple clustering between $D_m^+$ and $(O_2)_n$ as the source of their high abundance, or 'magic' character. Instead we attribute their high abundance to ion-molecule reactions which deliver specific ionic products with significant stabilities (see below). Clearly one option is deuteron transfer by reaction of $D_3^+$ with the oxygen dimer, $(O_2)_2$, although larger oxygen cluster may also contribute to the $DO_4^+$ signal. For ions containing $O_2$ the only ion with magic character is $D_2O_2^+$, which shows a very prominent excess abundance. For $O_6$ the most strongly magic ion is clearly $D_2O_6^+$, although $DO_6^+$ also shows significant abundance. For $O_{12}$ we see the greatest abundance for $D_2O_{12}^+$ and $D_3O_{12}^+$ and this is typical for ions derived from even larger $(O_2)_n$ clusters (not shown here).

The marked difference between the ion yields for $O_2$ and $(O_2)_2$ is potentially revealing about the ion-molecule chemistry taking place in helium droplets. We assume that the formation of $D_mO_x^+$ ions can be initiated by reactions of $D_3^+$ or their cluster equivalents, $D_3^+(D_2)_p$, although for simplicity we will restrict discussion to the former. Presumably the production of $D_3^+$ is initiated by collision of the droplet with a 70 eV electron, which can generate either metastable electronically excited helium (He*) or $He^+$ in the droplet. These reagents then ionize $D_2$ either by Penning ionization (He*) or charge transfer ($He^+$). Formation of the lowest metastable state of atomic helium requires 20.6 eV of energy and the ionization threshold lies at 24.6 eV.[19] Since the adiabatic ionization energy of $D_2$ is at 15.4 eV,[20] ionization of $D_2$ via either route will deliver several eV of excess energy into the helium droplet. In principle this excess energy will appear as heat and is far in excess of that necessary to initiate the deuterated equivalent of reaction (1). In view of this the absence of any strongly abundant $DO_2^+$ product can be explained in two ways: (a) the ionized cluster aggregate is quickly cooled by the surrounding helium after $D_3^+$ is made and therefore reaction with $O_2$ is prevented by the small but non-zero endothermicity, or (b) a secondary reaction takes place that efficiently removes the $DO_2^+$. Explanation (a) is unlikely, since it has been shown in many previous studies of ion-molecule reactions in helium droplets that the products are often consistent with hot reaction conditions, despite the very low temperature and high thermal conductivity of superfluid helium. It seems that relatively slow reactions resulting from significant structural rearrangement can be quenched,[21] whereas simple bond fissions are often too fast for even superfluid helium to provide any quenching.[22-25]

For explanation (b) we presume that the reaction

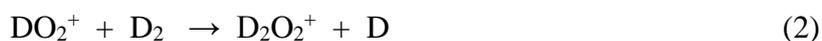

$$DO_2^+ \ + \ D_2 \ \rightarrow \ D_2O_2^+ \ + \ D \qquad\qquad (2)$$

takes place. Using the enthalpies of formation at 0 K of gaseous $HO_2^+$ (1109 kJ mol[-1])[26], H (216 kJ mol[-1])[27] and $H_2O_2$ (-130 kJ mol[-1])[26] together with the adiabatic ionization energy of $H_2O_2$ (1021 kJ mol[-1]),[28] and assuming that deuteration has a negligible effect on these thermodynamic quantities, we can calculate the enthalpy change for reaction (2). We find that the reaction is essentially thermoneutral, with an exothermicity of only 2 kJ mol[-1] and a margin of error of comparable size. Gas phase kinetic studies have shown that this reaction is close to the collision-limited rate[29] and so it is certainly plausible that reaction (2) could efficiently consume any $DO_2^+$ formed.

The reaction between $(O_2)_2$ and $D_3^+$ has a different end-product distribution, with the dominant ion being $DO_4^+$. In this case there is no doubt that proton transfer from $D_3^+$ to $(O_2)_2$ should readily occur because the reaction is exothermic and this is consistent with the observed ion abundance. However, the experimental data also indicate that $DO_4^+$ is much less willing than $DO_2^+$ to undergo a secondary reaction with $D_2$. In Figure 3 we further illustrate the predominance of $DO_4^+$ production by showing a plot of the ratio of $DO_x^+$ to the $D_2O_x^+$ and $D_3O_x^+$ ion signals for ions with even $x$ in the range $2 \leq x \leq 12$. We suspect that the $D_3O_x^+$ ions, which become the most abundant ions for large $(O_2)_n$ clusters, are derived from the secondary association reaction between $DO_x^+$ and $D_2$. Figure 3 shows that $DO_4^+$ is by far the most resistant of the $DO_x^+$ ions to secondary reactions, suggesting an enhanced stability for $DO_4^+$.

Although we have provided experimental evidence which supports a theoretical prediction that $HO_4^+$ can form from reactions at low temperature,[10] and that this ion is stable, the possibility of using $HO_4^+$ as a tracer molecule for $O_2$ in the ISM is questionable. The principal obstacle here is the formation of the $(O_2)_2$ dimer,.[13] It is not clear where the three-body collisions necessary to form this dimer could come from in the highly dilute conditions of the ISM. However, we note that there are other sources of oxygen in astronomical environments from which oxygen clusters might be formed. For example, Bieler *et al.* have recently reported a surprisingly (several per cent) high content of molecular oxygen in the nucleus of a comet.[30] If oxygen is trapped in any significant quantities on cold grains and within water ice then release of dimers might be possible. In order to facilitate a possible search for $HO_4^+$, the infrared spectrum of this ion was recently predicted from *ab initio* calculations.[11] There would certainly be value in carrying out further laboratory studies to characterize $HO_4^+$, and in particular to confirm its spectroscopic signature.

**Acknowledgements**

This work was given financial support by the Austrian Science Fund (FWF) Wien (P26635 and I978).

**Figure captions**

1.      Part of the mass spectrum from electron ionization of helium nanodroplets containing a $D_2/O_2$ mixture. In the main image peaks from the $O_6^+$ and $O_8^+$ ions are highlighted. Built upon the $O_6^+$ and $O_8^+$ peaks are additional peaks from ions with added D atoms. Those ions with an even number of D atoms have masses which notionally coincide with $He_n^+$ cluster ion peaks. However, the mass resolution is high enough to distinguish between these peaks. The high mass resolution is demonstrated in the inset. A deconvolution leads to the dashed lines in the inset, which show underlying contributions from ions with particular values of $m/z$.

2.      Plot of the signal for $D_mO_x^+$ ions versus $m$ for $x = 2$, 4, 6 and 12.

3.      Plot of the ion signal ratios $DO_x^+/D_2O_x^+$ and $DO_x^+/D_3O_x^+$ versus the number of oxygen atoms, $x$.

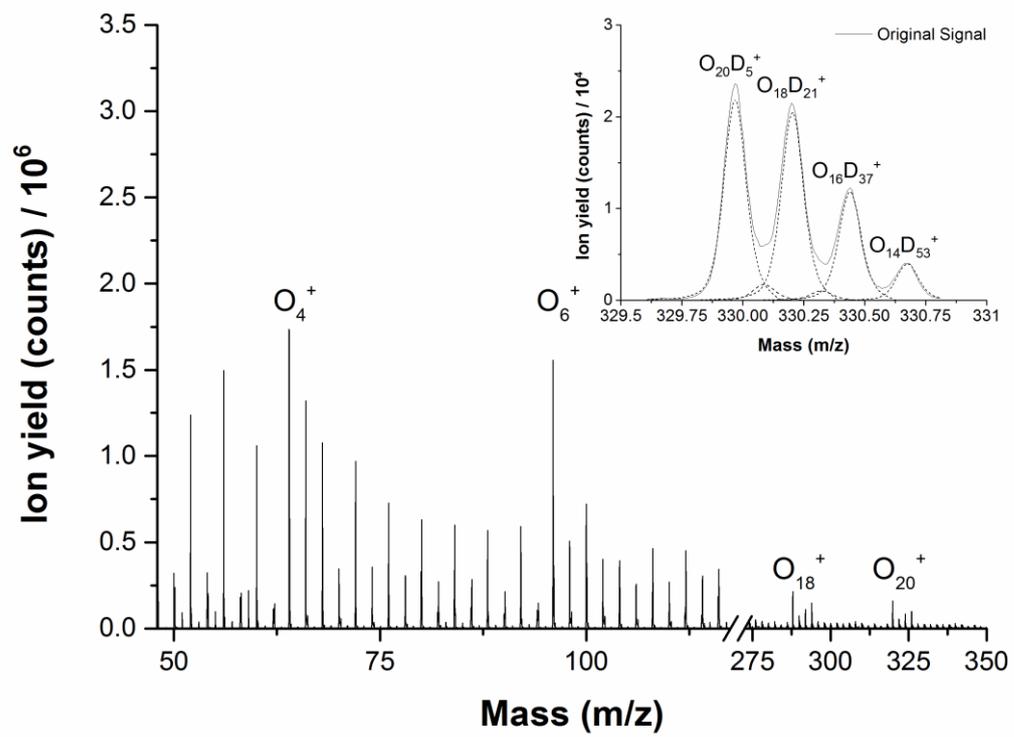

Figure 1

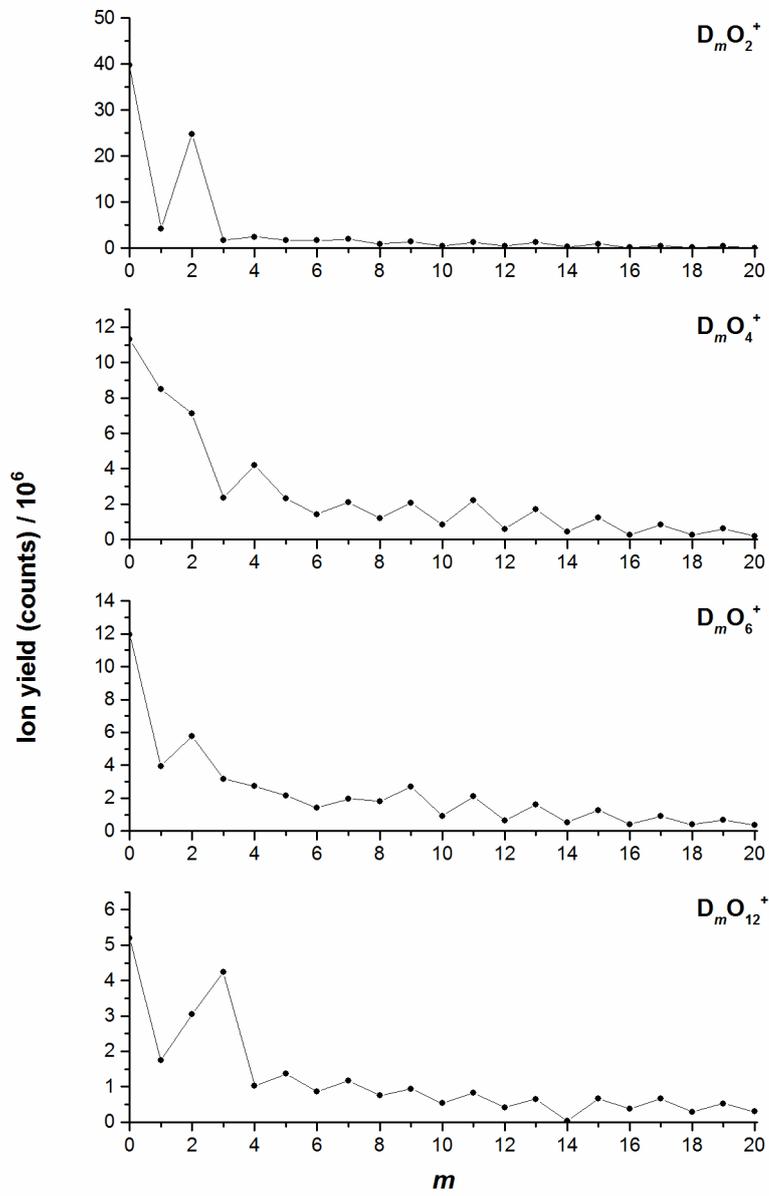

Figure 2

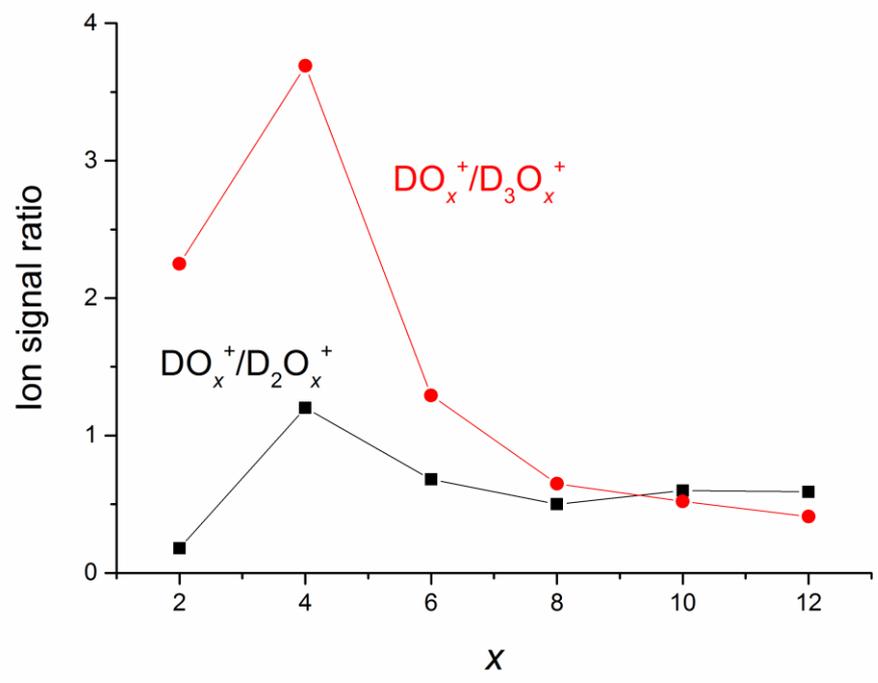

Figure 3